\documentclass[]{spie}  

\usepackage{amsmath,amsfonts,amssymb}
\usepackage{graphicx}
\usepackage[colorlinks=true, allcolors=blue]{hyperref}

\usepackage{amsmath,amsfonts,bm}

\def\eqref#1{equation~\ref{#1}}

\def\1{\bm{1}}

\def\vw{{\bm{w}}}

\def\vy{{\bm{y}}}

\def\mX{{\bm{X}}}
\def\mY{{\bm{Y}}}
\def\mZ{{\bm{Z}}}

\DeclareMathAlphabet{\mathsfit}{\encodingdefault}{\sfdefault}{m}{sl}
\SetMathAlphabet{\mathsfit}{bold}{\encodingdefault}{\sfdefault}{bx}{n}
\newcommand{\tens}[1]{\bm{\mathsfit{#1}}}

\def\tX{{\tens{X}}}

\def\sC{{\mathbb{C}}}

\def\sT{{\mathbb{T}}}

\newcommand{\E}{\mathbb{E}}

\usepackage{hyperref}
\usepackage{url}
\usepackage{xcolor}
\usepackage{graphicx}
\graphicspath{ {./images/} }

\newcommand*\mean[1]{\bar{#1}}

\title{Explaining neural network predictions of material strength}

\author[a,*]{Ian A. Palmer}
\author[a,*]{T. Nathan Mundhenk}
\author[a]{Brian Gallagher}
\author[a]{Yong Han}
\affil[a]{Lawrence Livermore National Laboratory, Livermore, CA, USA}
\affil[*]{Equal Contribution}

\authorinfo{Send machine learning correspondence to T. Nathan Mundhenk or materials questions to Yong Han\\E-mail: T. Nathan Mundhenk, mundhenk1@llnl.gov\\E-mail: Yong Han, han5@llnl.gov}

\begin{document} 
\maketitle

\begin{abstract}
We recently developed a deep learning method that can determine the critical peak stress of a material by looking at scanning electron microscope (SEM) images of the material’s crystals. However, it has been somewhat unclear what kind of image features the network is keying off of when it makes its prediction. It is common in computer vision to employ an explainable AI saliency map to tell one what parts of an image are important to the network’s decision. One can usually deduce the important features by looking at these salient locations. However, SEM images of crystals are more abstract to the human observer than natural image photographs. As a result, it is not easy to tell what features are important at the locations which are most salient. To solve this, we developed a method that helps us map features from important locations in SEM images to non-abstract textures that are easier to interpret.
\end{abstract}


\section{Introduction}

\begin{figure}[h]
\begin{center}
\includegraphics[width=0.8\linewidth]{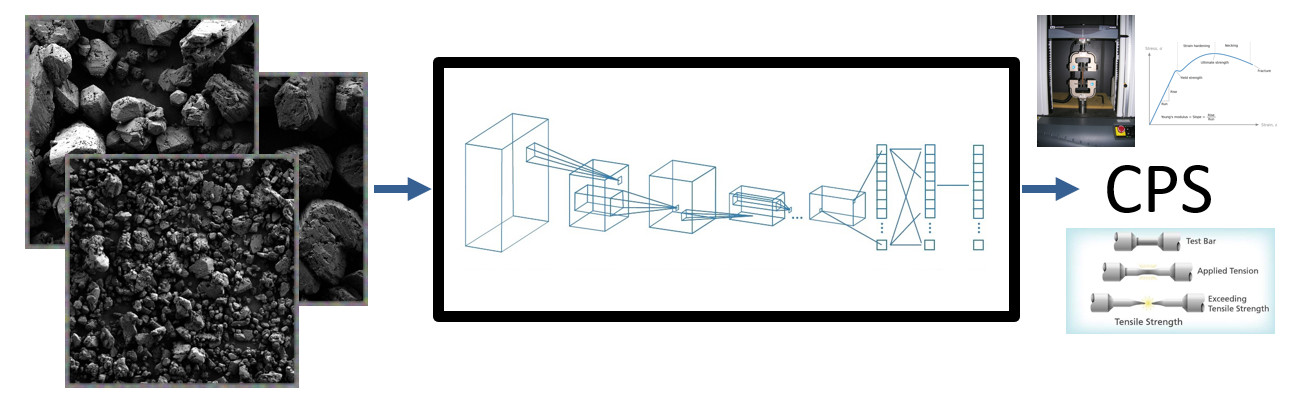}
\end{center}
\caption{SEM images are taken of \emph{Triaminotrinitrobenzene} (TATB) crystals produced in different lots. The production process is varied for each lot which yields TATB with different properties. A \emph{convolutional neural network} (CNN) is trained to predict the CPS based on the known ground truth obtained from physical stress strain tests.}
\label{fig:cps}
\end{figure}

Traditional materials engineering processes are iterative and include repeated cycles of destructive testing.  In testing a material's tensile strength, for example, a sample must be pulled apart until the material irreversibly deforms.  For many materials of interest, this is expensive and potentially dangerous. Figure \ref{fig:cps} shows how neural networks have been shown to be capable of predicting a material's \emph{critical peak stress} (CPS) given only \emph{Scanning Electron Microscope} (SEM) images \cite{gallagher01}.  This reduces the cost of producing materials for destructive testing and is significantly faster and cheaper than traditional methods.  However, a CPS prediction alone is of limited use - materials engineers require additional information about how that prediction is being made so they can iterate on the composition of the material.  Explainable AI techniques offer one solution to this problem. 

Prior work in the field of explainable AI has focused largely on generating visual representations of a CNN's attention field \cite{selvaraju01, chattopadhyay01, mundhenk01}.  These techniques are useful for confirming that a model has correctly learned to associate a label with the occurrence of that concept in an input image.  In many domains, however, a saliency map alone is not enough to interpret \emph{what} a neural network is \emph{"seeing"} in the input image.  This is particularly true for non-natural images, like SEM imagery. For example, a saliency map might highlight the eyes of a dog and its ears in a photograph. We might then conclude that the eyes and ears are important for the neural network. Images of crystals lack these obvious semantic components and are harder to interpret from saliency maps alone. This is also important since in this domain, analysts are primarily interested in what features are driving a neural network's prediction, not just the location in the image where those features occur.  This makes previous explainable AI work insufficient at answering how models predict CPS in SEM images.

Much of this work is built on top of \emph{FastCAM} \cite{mundhenk01}.  FastCAM is a saliency map technique that is computationally more efficient than other methods like \emph{Guided Backprop} \cite{Springenberg15}, \emph{Integrated Gradients} \cite{Sundararajan17}, \emph{Full Grad} \cite{Srinivas19} and \emph{Smooth Grad} \cite{Smilkov17} but at the same time has better fidelity.  It achieves this by taking advantage of Saliency Order Map Equivalence (SMOE), which allows saliency maps at different spatial resolutions to be combined in a computationally inexpensive manner.  While we chose to use Fast-CAM to compute saliency maps over images, this framework is in theory extensible to any similar explainable AI technique.

Our technique begins with a CNN and the dataset on which that model is trained.  In addition, we use another dataset whose purpose is to provide classes and images that are more easily interpretable by a human than the original dataset.  In the use case explored in this paper, our original dataset is a collection of SEM images of materials and the interpretable dataset is called the \emph{Describable Textures Dataset} (DTD) \cite{cimpoi01}.  First, a forward pass through the model with Fast-CAM is computed for each image in both the model's trained SEM dataset and the interpretable DTD dataset.  We then cluster all of the layer activations in high-dimensional space.  For any given image or class in the original SEM training dataset, the nearest DTD images or classes can be extracted from the shared feature space.  This allows our technique to provide both qualitative and quantitative answers to the question "What textures in this SEM image is the model using to predict CPS?".

\section{Related Work}

Some existing work has been done on the application of layer activations to measuring texture similarity.  Gao \emph{et al.} \cite{gao01} use CNN layer activations to estimate the similarity between pairs of textures and compare the predicted similarity values with data collected from human perception.  They use cosine similarity to compare layer activations for each pair of texture images, then feed those calculated similarity vectors into a fully-connected layer.  This approach has some fundamental similarities to our work, as they make the same basic hypothesis that the layer activations of a CNN contain information that can be used to identify how similar two textures are.  They also use the cosine similarity of layer activation vectors.  However, their work is primarily directed towards predicting human perceptual behavior, not analyzing other natural images.

Girish \emph{et al.} \cite{girish01} use k-means to cluster layer activations.  For every image in their query set, they compute the activations for every layer in the CNN.  Separated by layer, they then cluster those activations and retrieve the images that belong to certain clusters.  Based on their findings, earlier layers learn lower-level features and later layers learn features that distinguish image classes.  Their findings are consistent with our interpretation of the features at different layers in a CNN, but while their work focuses on clustering natural images from a similar distribution, we compare layer activations between both natural images and texture images.

\section{Methodology}

\subsection{Interpretable Dataset}
\begin{figure}[h]
\begin{center}
\includegraphics[width=0.8\linewidth]{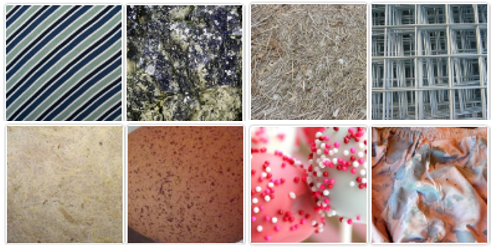}
\end{center}
\caption{Sample images from the Describable Textures Dataset from many different texture classes are shown.}
\label{fig:dtd}
\end{figure}

Fundamentally, the idea of this technique is to analyze a model's behavior when presented with difficult-to-interpret data by comparing it with its behavior when presented with easy-to-interpret data.  As a result, the choice of comparison dataset is a determining factor in the results of this analysis.  Our inspiration to select a texture-based dataset came from our interest in what was driving the model to select certain regions of the SEM images as salient (or not) and our hypothesis that it was the structure of the materials in question.  The Describable Textures Dataset, with 47 classes and 120 samples per class, provides appropriate levels of class diversity and intra-class variance.  Figure \ref{fig:dtd} shows several samples from this dataset.

\subsection{SEM Dataset}

The subjects of our SEM image dataset are crystals of TATB.  Each of the 30 different alphabetically labeled batches of TATB was synthesized under different conditions, approximately covering the known feature space of the material \cite{gallagher01}.  

In total, the dataset is comprised of 69,894 samples evenly distributed across the different batches.  Images are scaled to 384 by 384 pixels via bilinear interpolation and converted to RGB from grayscale.  The images are then normalized and randomly cropped to 352 by 352 pixels.  For our purposes, every sample is associated with the batch's experimentally measured critical peak stress value.

\subsection{Model Training}
In this experiment, we used a ResNet-50 \cite{he01} architecture.  We started with a model pretrained on ImageNet \cite{deng01}.  The training objective was to predict the batch's CPS given a sample from that batch via regression.  We trained the models using an initial learning rate of 0.01 and exponentially decreased the rate.  We used a batch size of 32.  Stochastic gradient descent was used as an optimizer with a momentum of 0.9 and weight decay of 0.0002.  We trained the model until the loss stopped decreasing over multiple epochs, then chose the model with the best validation score.  Overall, our approach was similar to that of Gallagher \emph{et al.} \cite{gallagher01}, but we use a ResNet-50 instead of a DenseNet \cite{DenseNet} since it gives slightly better CPS predictions.

\subsection{Feature Generation}
\label{sec:feature-gen}
Feature generation begins with a ResNet50 \cite{he01} convolutional neural network $\displaystyle \mZ$, comprised of 5 convolutional stages.  $\displaystyle \mZ$ has weight parameters $\theta$ and acts on an argument $\displaystyle \tX$ to produce a single real-valued CPS prediction $\displaystyle \mZ \left( \displaystyle \tX ; \theta \right) = \displaystyle \mY$.  For each of the $47 \times 5 = 5640$ samples $\displaystyle \tX$ in DTD, we compute $\displaystyle \mY$.  Next, we extract the outputs of the first max pooling layer and each of the four identity blocks.  We label these $\displaystyle \mZ_1$ through $\displaystyle \mZ_5$. The activations are feature vectors extracted at the most salient spatial location in each block. These activations are saved to disk along with the saliency map generated by Fast-CAM in order to reduce repeated computation.

Overall, this is a fairly fast process.  DTD is smaller than most contemporary machine learning datasets, and only a forward pass is required to extract the layer activations and Fast-CAM maps.  Running on a GPU-accelerated system, this process generally took less than an hour.

\section{Experiments}

\subsection{Sample Clustering}
\label{sec:sample-clustering}

The most straightforward analysis that can be performed on these layer activations is a comparison between data from the model's trained distribution (SEM images in this case) and the saved layer activations.  In this mode, one sample $\displaystyle \mX_{SEM}$ is passed through the network in the same manner as described in \ref{sec:feature-gen}.  The layer activations from this sample can then be projected into the same high-order feature space as the layer activations from the DTD samples.  The dimensionality of this feature space depends on the layer that activations are selected from, which is an adjustable parameter.  In our experiments, we found that the best results came from $\displaystyle \mZ_5$ (the output of the final identity block).

The $l^2$ norm is then calculated for the difference between each $\left( \displaystyle \mZ_{5, SEM}, \displaystyle \mZ_{5, DTD} \right)$ pair.  The distance metric $d$ is defined as follows:

\begin{equation}
     d \left( \displaystyle \mZ_1, \displaystyle \mZ_2 \right) = \sqrt{\sum_{k=1}^{n} \left( \displaystyle \mZ_1 - \displaystyle \mZ_2 \right)^2 }
\end{equation}
\label{eq:distance}

The $k$ nearest neighbors to $\displaystyle \mX_{SEM}$ in the high-dimensional space can then be extracted by selecting the $k$ smallest $d$ values and retrieving the corresponding $\displaystyle \mX_{DTD}$.

\begin{figure}[h]
\begin{center}
\includegraphics[width=0.8\linewidth]{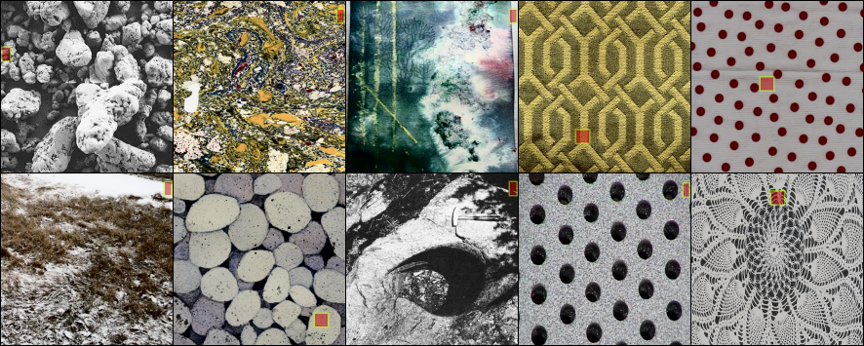}
\end{center}
\caption{The $k = 9$ nearest neighbors to an SEM sample (top left) based on layer activations.  The nearest neighbors are sorted in left-right then top-bottom order.  The highlighted regions are the most salient regions in each image according to the Fast-CAM saliency map.}
\label{fig:sample-sim}
\end{figure}

Figure \ref{fig:sample-sim} shows an example of the nearest neighbors to one SEM image sample.  The repeated occurrence of images from certain texture classes (\textit{dotted} and \textit{flecked} here, for example) suggests that these textures are more relevant to the model in predicting critical peak stress than other textures which do not appear in the nearest neighbors.

This approach has several limitations, though.  First, the results are more variable when the nearest neighbors of only one SEM image are extracted.  Additionally, it can be difficult to see from the texture images what exactly the model could be looking at.  To address these concerns, we performed additional experiments with a slightly varied technique.

\subsection{Class-Averaged Clustering}

Given the variance that can result from analyzing a single image, we conducted an additional set of experiments where distance values were averaged across all of the samples within a target SEM class.  While this increases the amount of computation required, it also provides more insight about which images in particular are most relevant to a class of interest.  For example, some classes may be suspected to contain more microscopic porosity than other classes.  Observing a large number of similarly porous or spotted textures in the retrieved neighbors is a qualitative measure, but it can provide evidence to support that conclusion.

The precise formulation of our class-averaged distance metric given a class $\displaystyle \sC = \{ \displaystyle \mX_1, \displaystyle \mX_2, \ldots, \displaystyle \mX_n \}$ is the following:

\begin{equation}
    \mean{d} \left( \displaystyle \sC, \displaystyle \mZ \right) = \frac{1}{n} \sum_{k=1}^{n} d \left(\displaystyle \sC_k, \displaystyle \mZ \right)
\end{equation}
\label{eq:mean-distance}

\begin{figure}[h]
\begin{center}
\includegraphics[width=0.8\linewidth]{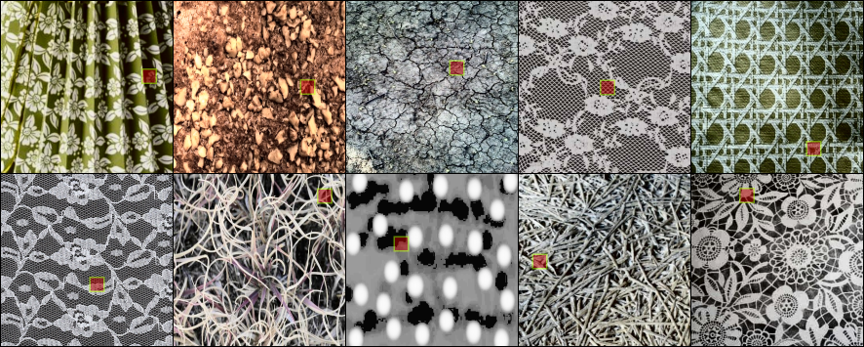}
\end{center}
\caption{The $k = 10$ nearest neighbors across a class of SEM images, sorted by increasing distance from left to right then top to bottom. Again, the highlighted regions are the most salient regions in each image according to the Fast-CAM saliency map. }
\label{fig:batch-sim}
\end{figure}

The results in Figure \ref{fig:batch-sim} are similar to those in Figure \ref{fig:sample-sim}, except now the distance metrics are averaged across all SEM images in the class.  This makes the retrieved texture images more resistant to noise in an individual SEM sample.  Figure \ref{fig:batch-sim} contains a number of texture samples from the \textit{lacelike} class, for example, suggesting that texture is particularly similar to what the model is looking at in the SEM image.

\subsection{Critical Peak Stress Correlation}
\label{sec:cpscorr}

While the qualitative methods described above can be useful to gain a visual understanding of what textures the model is looking at in the SEM images, it lacks a quantitative aspect and is dependent on the user having some knowledge of the domain in order to make informed interpretations.  Additionally, determining the textures that are most present in an image is less important than determining the textures that best correlate with a strong or weak material as measured by CPS.  As a result, we extend the previous techniques to generate quantitative rankings of the textures that have the highest and lowest correlations with CPS.

We achieve this by first constructing a vector with each class' real-valued critical peak stress.  These values are normalized to have $\mu = 0$ and $\sigma = 1$.  We then need to extract the relevance of each texture class to each SEM class.  To do this, the mean distance metric given by Equation \ref{eq:mean-distance} is averaged across all texture samples in a single class to give one real value $r$ representing the relevance of that texture to a SEM class.  This is shown in Equation \ref{eq:class-distance}, where $\displaystyle \sT = \{ \displaystyle \mX_1, \displaystyle \mX_2, \ldots, \displaystyle \mX_m \}$ is a class of textures.

\begin{equation}
    r \left( \displaystyle \sC, \displaystyle \sT \right) = \frac{1}{m} \sum_{k=1}^{m} \mean{d} \left(\displaystyle \sC, \displaystyle \sT_m \right)
\end{equation}
\label{eq:class-distance}

Once $r$ values have been obtained for each pair of $\displaystyle \sC$ and $\displaystyle \sT$, those values are arranged into $| \displaystyle \sC |$ vectors of length $| \displaystyle \sT |$.  These vectors match the size of the previously constructed vector of CPS values.  Like the values in that vector, the values in these vectors are normalized to have $\mu = 0$ and $\sigma = 1$.

Given these vectors, we are interested in how well each correlates with CPS values.  A high correlation between those vectors would indicate that the presence of that feature is correlated with the material having a higher strength, while a low correlation would imply that the presence of that texture in a crystal is correlated with the material having a low CPS.  No correlation between the two vectors would mean that the presence or absence of that texture in the crystals is not particularly indicative of overall material strength.

We use cosine similarity to measure the correlation between vectors.  Given the vector of CPS values $\displaystyle \vw$ and a vector of texture similarities $\displaystyle \vy$, the cosine similarity $s$ is given by the following:

\begin{equation}
    s = \frac{\displaystyle \vw \cdot \displaystyle \vy}{|| \displaystyle \vw || \cdot ||\displaystyle \vy ||}
\end{equation}
\label{eq:cosine}

\begin{figure}[h]
\begin{center}
\includegraphics[width=\linewidth]{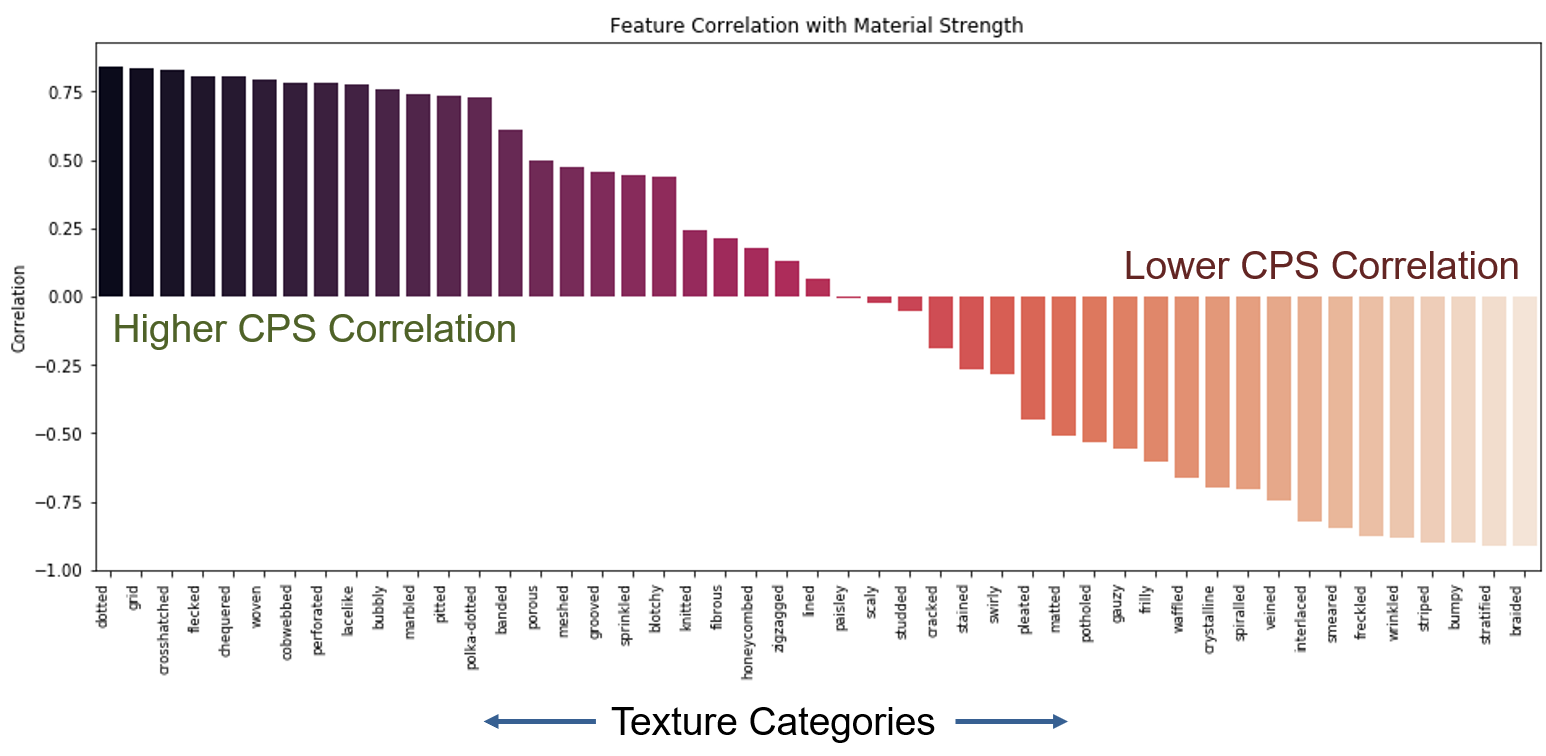}
\end{center}
\caption{Cosine similarity values for each texture class, showing how the presence of that texture (on average) influences the CPS of the material.}
\label{fig:cosine-1}
\end{figure}

A chart of each texture class' $s$ value is shown in Figure \ref{fig:cosine-1}.

\begin{figure}[h]
\begin{center}
\includegraphics[width=0.9\linewidth]{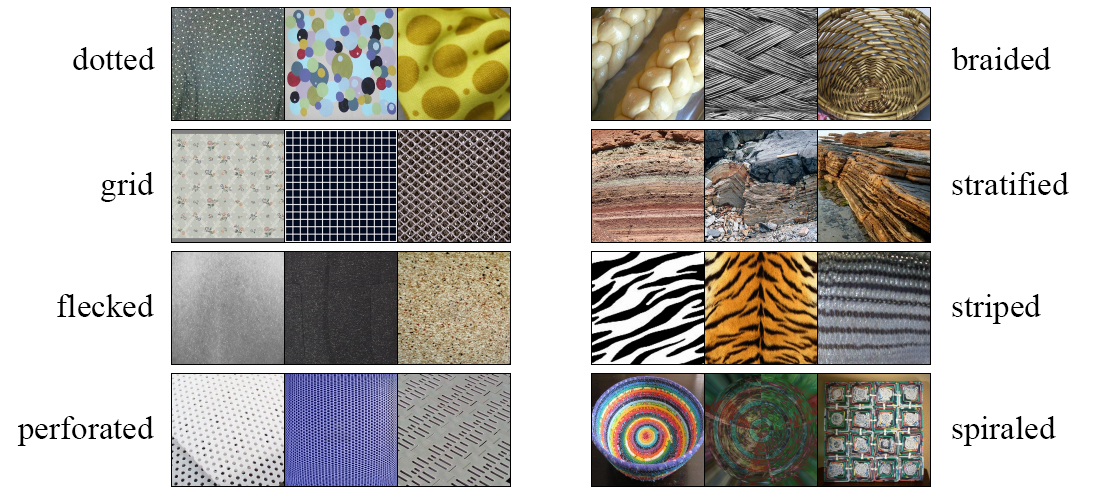}
\end{center}
\caption{Samples from the Describable Textures Dataset.  The classes on the left have a positive correlation with critical peak stress, while those on the right have a negative correlation with CPS.}
\label{fig:texture-sample}
\end{figure}

Figure \ref{fig:cosine-1} shows the results of our analysis using cosine similarity, and the same results are shown in tabular form in Figure \ref{tab:cosine-1}..  The presence of textures like \verb|dotted|, \verb|grid|, and \verb|flecked| is highly correlated with a material's experimentally determined CPS.  Examples of these textures are shown in Figure \ref{fig:texture-sample}.

\begin{table}[t]
\caption{Correlation with CPS}
\label{tab:cosine-1}
\begin{center}
\begin{tabular}{|p{3cm} c | p{3cm} c|}
\hline & & &  \\
\multicolumn{2}{|c|}{\bf Top 10} & \multicolumn{2}{|c|}{\bf Bottom 10} \\
\multicolumn{1}{|l}{\bf Texture}  &\multicolumn{1}{c|}{\bf Correlation} &
\multicolumn{1}{|l}{\bf Texture}  &\multicolumn{1}{c|}{\bf Correlation}
\\ \hline & & &  \\
dotted                  &$0.8421$  & braided            &$-0.9132$ \\
grid                    &$0.8324$ & stratified          &$-0.9131$ \\
crosshatched            &$0.8297$ & bumpy               &$-0.9004$ \\
flecked                 &$0.8079$ & wrinkled            &$-0.9001$ \\
chequered               &$0.8076$ & freckled            &$-0.8796$ \\
woven                   &$0.7947$ & smeared             &$-0.8762$ \\
cobwebbed               &$0.7816$ & interlaced          &$-0.8469$ \\
perforated              &$0.7804$ & veined              &$-0.8250$ \\
lacelike                &$0.7773$ & spiralled           &$-0.7439$ \\
bubbly                  &$0.7566$ & crystalline         &$-0.7007$ \\
 \hline
\end{tabular}
\end{center}
\end{table}

\section{Analysis}

\subsection{Sample Clustering}

Our sample clustering results are the \textit{qualitative} component of our analysis framework.  Primarily, these results show what textures in the SEM image are most salient to the prediction model.  Looking at the nearest neighbors of an individual image is useful especially in the case where the model makes an incorrect prediction of the material's CPS.  Our technique can shed light on the cause: for example, a sample from one class of material may show very different salient textures than other samples from the same class.  This may suggest that the region imaged in the outlier sample is abnormal compared to most of the SEM images belonging to that class.

We have also found that providing the qualitative results to the engineers creating the materials is useful.  Visually seeing some of the textures present in the SEM images can inform the practices and procedures used to create future batches of materials.

The class-averaged results are useful primarily in comparing the textures present in different batches.  Just as the nearest neighbor results for an individual image can be used to determine what the model sees in that particular sample, class-averaged results allow users to infer differences across classes.  Observing that a particular class of samples is relevant to one class of SEM images but not another can inform the measured differences in material properties (such as CPS). 

\subsection{Critical Peak Stress Correlation}
\label{sec:cps-analysis}

\begin{figure}[h]
\begin{center}
\includegraphics[height=8.0 cm]{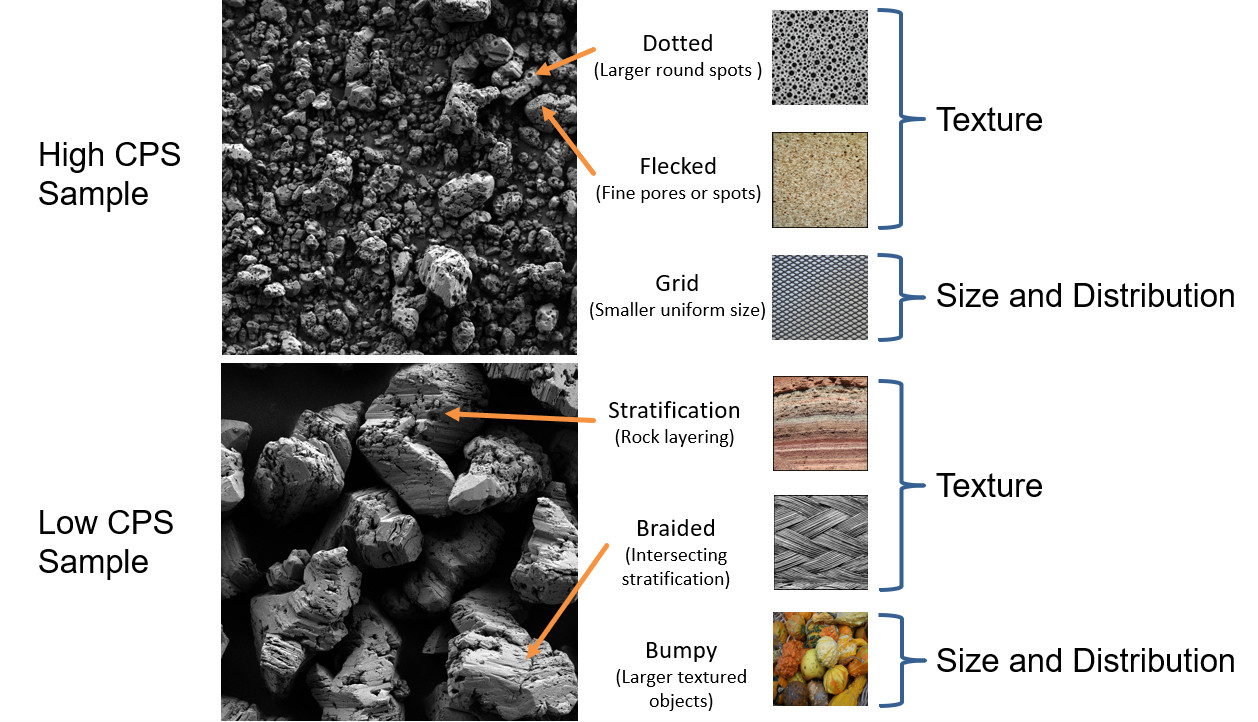}
\end{center}
\caption{The preponderance of many of the best correlating textures is easy to see in the crystals. However, some textures such as grid and bumpy are harder to see. We believe that these relate to size and distribution of crystals.}
\label{fig:interp}
\end{figure}

The results of our CPS analysis, summarized in Table \ref{tab:cosine-1} and Figure \ref{fig:cosine-1}, are particularly encouraging because they confirm previously held hypotheses about the microscopic mechanisms influencing CPS in these crystal materials.  The material of interest in the SEM images belongs to the family of rhombohedral crystals, and the crystals are bound together with a glue bonding agent.  We therefore hypothesize that the positive correlation of \textit{dotted}, \textit{flecked}, and \textit{perforated} textures (among others) with CPS is due to the ability of the bonding agent to adhere to a greater surface area in crystals with more microscopic pitting. Why the \textit{grid} class correlates well with CPS is less obvious, but the best correlating grids have a very similar size and distribution to high CPS crystals. As such, we suspect that grid class describes these two attributes. Examples of this can be seen in figure \ref{fig:interp}.  


The textures that tend to have a negative correlation with CPS, like \textit{braided}, \textit{stratified}, and \textit{striped} are fairly similar to each other.  Each contains long lines that extend from one end of the image to the other.  On a microscopic scale, we hypothesize that the model is picking up on cracks and similar deformities in the material.  These small fractures in the crystalline structure are likely to have a negative effect on the measured CPS of the material. It may also be that crystals are weak since the stratified layering might be pulled apart like shale. Similar to the higher CPS correlations, some are not as obvious. Like with grid, the best correlating \textit{bumpy} images seem to have a similar size and distribution as low CPS crystals.    

Finally, the textures that don't correlate particularly positively or negatively with CPS (those in the middle of Figure \ref{fig:cosine-1}) are those that we hypothesize are either not present in the materials (\textit{blotchy} and \textit{paisley}, for example) or present in essentially every sample (\textit{studded} and \textit{cracked}).  Others may also be present to different degrees in different classes but have no correlation to CPS.  This hypothesis is supported by our experiments in Section \ref{sec:sample-clustering}.  We saw that some textures are found to be similar to essentially every texture, while some textures are not found to be particularly similar to any class. That is, some textures are globally similar to all SEM feature vectors, regardless of CPS. Still, others show variance across classes but don't vary with CPS.  This phenomenon is explored in greater depth in the Appendix.\ref{fig:interp}

\section{Conclusion}

In this paper, we introduce a novel technique for interpreting convolutional neural networks by generating saliency maps and clustering both textures and images in a shared space.  Our approach provides insight into the prediction process of a CNN, and most importantly that insight is grounded in human-interpretable natural images.  Unlike some explainable AI techniques, the textures identified as being positively or negatively correlated with critical peak stress are interpretable by researchers that are less familiar with CNNs and saliency maps.  This makes the process useful for materials researchers and engineers, who can then rapidly iterate on the design of the target material to optimize its properties.  In practice, we have found that this technique is a useful addition to the materials engineering pipeline and has the potential to save money and effort that would otherwise be spent on expensive destructive testing.

\subsection{Future Work}

Future work on this explainable AI technique could focus on the use of a larger comparison dataset (DTD in this work).  That dataset could be expanded to include natural images, which are often rich in content and could provide interesting insights into what information the CNN is using in the target images to predict their properties.

Additionally, this technique could be applied to other tasks.  While our task was a regression task from images to numerical values, this technique may work on other tasks like object detection, classification, or localization.  The versatility of CNNs suggests this technique may also have wide-reaching applications.

\subsubsection*{Acknowledgments}
This work was performed under the auspices of the U.S. Department of Energy by Lawrence Livermore National Laboratory under Contract DE-AC52-07NA27344 and was supported by the LLNL-LDRD Program under projects 19-SI-001, 18-ERD-021 and 17-SI-003.



\bibliographystyle{spiebib} 
\bibliography{iclr2020_conference}

\begin{thebibliography}{10}

\bibitem{gallagher01}
Gallagher, B., Rever, M., Loveland, D., Mundhenk, T.~N., Beauchamp, B.,
  Robertson, E., Jaman, G.~G., Hiszpanski, A.~M., and Han, Y.-J., ``Predicting
  compressive strength of consolidated molecular solids using computer vision
  and deep learning,'' {\em Materials \& Design}~{\bf 190} (2020).

\bibitem{selvaraju01}
Selvaraju, R.~R., Cogswell, M., Das, A., Vedantam, R., Parikh, D., and Batra,
  D., ``Grad-cam: Visual explanations from deep networks via gradient-based
  localization,'' {\em IJCV}  (2019).

\bibitem{chattopadhyay01}
Chattopadhyay, A., Sarkar, A., Howlader, P., and Balasubramanian, V.~N.,
  ``Grad-cam++: Improved visual explanations for deep convolutional networks,''
  {\em WACV}  (2018).

\bibitem{mundhenk01}
Mundhenk, T.~N., Chen, B.~Y., and Friedland, G., ``Efficient saliency maps for
  explainable ai,'' {\em arXiv:1911.11293}  (2020).

\bibitem{Springenberg15}
Springenberg, J.~T., Dosovitskiy, A., Brox, T., and Riedmille, M., ``Striving
  for simplicity: The all convolutional net.,'' in [{\em ICLR
  Workshop}{\nolinebreak\hspace{0.1em}]},  (2015).

\bibitem{Sundararajan17}
Sundararajan, M., Taly, A., and Yan, Q., ``Axiomatic attribution for deep
  networks.,'' in [{\em ICML}{\nolinebreak\hspace{0.1em}]},  (2017).

\bibitem{Srinivas19}
Srinivas, S. and Fleuret, F., ``Full-gradient representation for neural network
  visualization.,'' in [{\em NIPS}{\nolinebreak\hspace{0.1em}]},  (2019).

\bibitem{Smilkov17}
Smilkov, D., Thorat, N., Kim, B., Viégas, F., and Wattenberg, M.,
  ``Smoothgrad: removing noise by adding noise.,'' in [{\em
  arXiv:1706.03825}{\nolinebreak\hspace{0.1em}]},  (2017).

\bibitem{cimpoi01}
Cimpoi, M., Maji, S., Kokkinos, I., Mohamed, S., , and Vedaldi, A.,
  ``Describing textures in the wild,'' in [{\em Proceedings of the {IEEE} Conf.
  on Computer Vision and Pattern Recognition
  ({CVPR})}{\nolinebreak\hspace{0.1em}]},  (2014).

\bibitem{gao01}
Gao, Y., Gan, Y., Qi, L., Zhou, H., Dong, X., and Dong, J., ``A
  perception-inspired deep learning framework for predicting perceptual texture
  similarity,'' ~{\bf 30} (2020).

\bibitem{girish01}
Girish, D., Singh, V., and Ralescu, A., ``Unsupervised clustering based
  understanding of cnn,'' {\em CVPR Explainable AI Workshop}  (2019).

\bibitem{he01}
He, K., Zhang, X., Ren, S., and Sun, J., ``Deep residual learning for image
  recognition,'' in [{\em Proceedings of the {IEEE} Conf. on Computer Vision
  and Pattern Recognition ({CVPR})}{\nolinebreak\hspace{0.1em}]},  (2016).

\bibitem{deng01}
Deng, J., Dong, W., Socher, R., Li, L.-J., Li, K., and Fei-Fei, L., ``Imagenet:
  A large-scale hierarchical image database,'' in [{\em Proceedings of the
  {IEEE} Conf. on Computer Vision and Pattern Recognition
  ({CVPR})}{\nolinebreak\hspace{0.1em}]},  (2009).

\bibitem{DenseNet}
Huang, G., Liu, Z., Weinberger, K.~Q., and van~der Maaten, L., ``Densely
  connected convolutional networks.,'' in [{\em
  CVPR}{\nolinebreak\hspace{0.1em}]},  (2017).

\end{thebibliography}

\pagebreak

\appendix
\section{Appendix}

\begin{figure}[htbp]
\begin{center}
\includegraphics[width=1.0\linewidth]{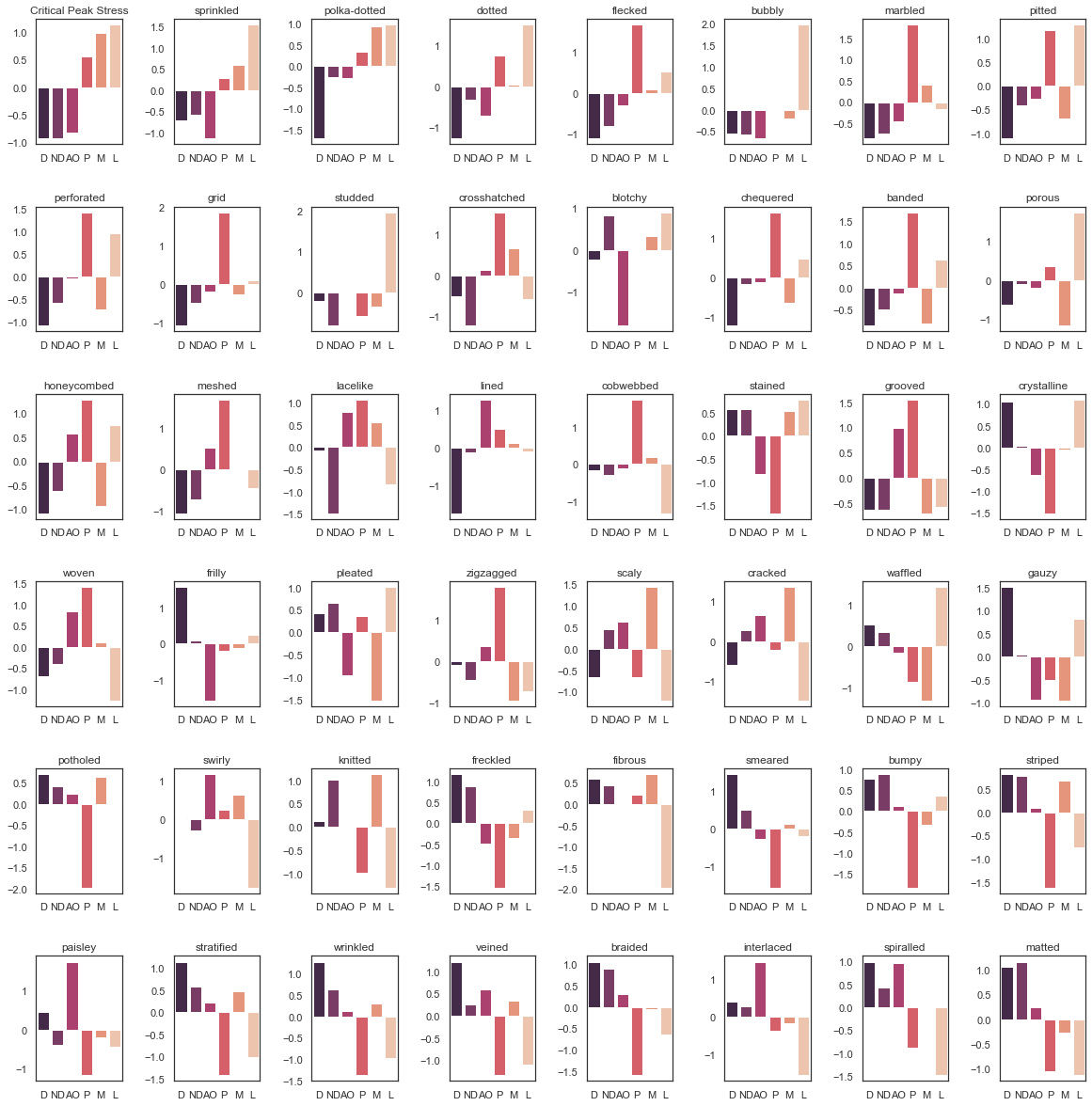}
\end{center}
\caption{Batch-wise correlations for a subset of 6 (of the 30 total) batches.  The top left chart shows how CPS varies between the batches, then the texture classes are sorted by decreasing correlation from top to bottom and left to right.  While this is based on a small subset of the total dataset, this is the same process described in \ref{sec:cpscorr}.  Some texture classes have a positive correlation to CPS, some have a negative correlation, and some exhibit an uncorrelated pattern.}
\label{fig:all-textures}
\end{figure}

Figure \ref{fig:all-textures} explores some of the correlations described in Section \ref{sec:cps-analysis}.  Some texture classes are positively correlated with CPS (top row), some are negatively correlated with CPS (bottom row) and some exhibit no particular correlation (middle row).  This data was extracted from a smaller subset of the larger SEM dataset, but the same general trends are visible here.  

\begin{figure}[htbp]
\begin{center}
\includegraphics[height=8.0 cm]{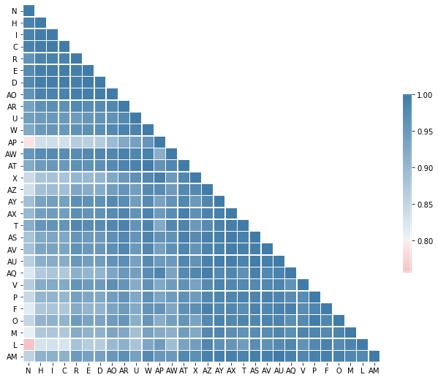}
\end{center}
\caption{This chart shows how similar the correlations were for each batch of SEM images.  For example, batches N and H had similar correlation values for each texture class, so they have a high similarity in this chart.  Similarly, batches N and L had different correlation values for each texture class, so they have a low similarity.}
\label{fig:batch-corr}
\end{figure}

Figure \ref{fig:batch-corr} shows how similar each batch is to every other batch.  Of note here is that the batches are organized by increasing CPS from left to right and top to bottom (so N has the smallest CPS and AM has the greatest CPS).  The pattern we observed is that batches N through AT (batches with a smaller than average CPS) tend to correlate highly with each other, while batches X though AM (batches with a larger than average CPS) tend to correlate highly with each other.  However, batches N-AT do not correlate well with batches X-AM.  This suggests that there is some fundamental textural difference that correlates with the experimentally measured CPS, supporting our finding that material texture plays a large role in the process of a CNN predicting critical peak stress values.

\end{document}